\documentclass[]{aa}
\usepackage{graphics,epsfig,txfonts}

\usepackage{natbib}
\bibpunct{(}{)}{;}{a}{}{,}

\newcommand{\expo}[1]{$\times 10^{#1}$}
\newcommand{\oexpo}[1]{$10^{#1}$}

\newcommand{\Halp}{H${\alpha}$}
\newcommand{\ltsima}{$\buildrel < \over \sim$}
\newcommand{\lsim}{\lower.5ex\hbox{\ltsima}}
\newcommand{\gtsima}{$\buildrel > \over \sim$}
\newcommand{\gsim}{\lower.5ex\hbox{\gtsima}}
\newcommand{\xmm}{XMM-{\it Newton}}
\newcommand{\cxo}{{\it Chandra}}
\newcommand{\swift}{{\it Swift}}

\newcommand{\sxp}{\hbox{SXP\,1062}}

\usepackage{color}

\begin{document}
 
\title{Long-term evolution of the neutron-star spin period of SXP\,1062
       \thanks{Based on observations obtained with XMM-Newton, an ESA science mission with instruments and contributions directly funded by ESA Member States and NASA
                 and on observations made with the Southern African Large Telescope (SALT)}
       \thanks{The reduced SALT spectra will be available at the CDS via anonymous ftp to cdsarc.u-strasbg.fr (130.79.128.5)
               or via http://cdsweb.u-strasbg.fr/cgi-bin/qcat?J/A+A/}
      }

\author{R.~Sturm\inst{1} 
   \and F.~Haberl\inst{1} 
   \and L.~M.~Oskinova\inst{2} 
   \and M.~P.~E.~Schurch\inst{3} 
   \and V.~H\'enault-Brunet\inst{4}    
   \and J.~S.~Gallagher III\inst{5}      
   \and A.~Udalski\inst{6}}

\titlerunning{Long-term evolution of the neutron-star spin period of SXP\,1062.}
\authorrunning{Sturm et al.}

\institute{ Max-Planck-Institut f\"ur extraterrestrische Physik, Giessenbachstra{\ss}e, 85748 Garching, Germany
	    \and
            Institute for Physics and Astronomy, University Potsdam, 14476 Potsdam, Germany
            \and
            Astrophysics, Cosmology and Gravity Centre (ACGC), Astronomy Department, University of Cape Town, Rondebosch, Private Bag X1 7701, South Africa
            \and
            Scottish Universities Physics Alliance (SUPA), Institute for Astronomy, University of Edinburgh, Blackford Hill, Edinburgh EH9 3HJ, UK
            \and
            Department of Astronomy, University of Wisconsin-Madison, 5534 Sterling, 475 North Charter Street, Madison, WI 53706, USA
            \and
            Warsaw University Observatory, Aleje Ujazdowskie 4, 00-478 Warsaw, Poland
	   }

\date{Received 22 April 2013 / Accepted 2 July 2013}

 \abstract{The Be/X-ray binary SXP\,1062 is of especial interest owing to the large spin period of the neutron star, its large spin-down rate, and the association with a supernova remnant constraining its age. 
          This makes the source an important probe for accretion physics.} 
          {To investigate the long-term evolution of the spin period and associated spectral variations, we performed an \xmm\ target-of-opportunity observation of SXP\,1062 during X-ray outburst.}
          {Spectral and timing analysis of the \xmm\ data was compared with previous studies, as well as complementary \swift/XRT monitoring and optical spectroscopy with the SALT telescope were obtained.}
          {The spin period was measured to be $P_{\rm s} =(1071.01\pm0.16)$ s on 2012 Oct 14. 
           The X-ray spectrum is similar to that of previous observations. No convincing cyclotron absorption features, which could be indicative for a high magnetic field strength, are found. 
           The high-resolution RGS spectra indicate the presence of emission lines, which may not completely be accounted for by the SNR emission. 
           The comparison of multi-epoch optical spectra suggest an increasing size or density of the decretion disc around the Be star.}
          {\sxp\ showed a net spin-down with an average of $\dot{P}_{\rm s} = (2.27\pm0.44)$ s yr$^{-1}$ over a baseline of 915 days.}

%


\maketitle

\section{Introduction}
\label{sec:introduction}
Be/X-ray binaries \citep[BeXRBs, ][]{2011Ap&SS.332....1R} are the dominant subclass of high-mass X-ray binaries in the Small Magellanic Cloud (SMC) with more than 100 known systems (including candidates).
These systems comprise of a neutron star (NS) that accretes matter from the circumstellar decretion disc of an emission-line star of spectral class B (Be star) leading to, often transient, X-ray emission.

The BeXRB binary \object{SXP\,1062} was discovered by \citet{2012MNRAS.420L..13H}
using \cxo\ and \xmm\ observations of the star-forming region NGC\,602 \citep{2013ApJ...765...73O}.
Amongst other BeXRBs in the SMC, \sxp\ is outstanding owing to: 
(i) The position in the wing of the SMC, where a younger stellar population is found and less BeXRBs are known compared to the bar of the SMC.
(ii) A relatively long spin period of the NS of $P_{\rm s}\sim1062$ s, making it the second slowest known pulsar of the SMC \citep[after SXP\,1323, ][]{2005A&A...438..211H}.
(iii) A rapid spin-down of the NS of $\dot{P}_{\rm s} = 0.26$ s d$^{-1}$ \citep{2012A&A...537L...1H} that was observed over the interval between 2010 Mar 25 and Apr 12 (18 days), 
whereas in general spin-up is observed during enhanced accretion.
(iv) A robust correlation with a supernova remnant (SNR) was found by \citet{2012MNRAS.420L..13H}.
Owing to the low density of both BeXRBs and SNRs in the wing of the SMC, this correlation is unlikely by chance.
The SNR is clearly seen in optical emission lines (\Halp\ and [\ion{O}{iii}]).
Additional radio emission and a detailed analysis of the X-ray emission of the SNR can be found in \citet{2012A&A...537L...1H}.
The SNR constrains the age of the NS to 10\,000$-$25\,000 years, making \sxp\ a relatively young BeXRB.

These peculiar properties have made \sxp\ the subject of several theoretical investigations, 
with the main issue of explaining the long spin period of this young NS.
Suggested scenarios include an initially slow spin period \citep{2012A&A...537L...1H},
an initially and/or presently strong magnetic field \citep{2012MNRAS.421L.127P,2012ApJ...757..171F},
and the accretion of magnetised matter \citep{2012MNRAS.424L..39I}.

Monitoring of Galactic accreting pulsars revealed that individual NSs can switch between spin-up and spin-down phases over short time scales down to a few days 
and that the NS spins up during intervals of high accretion and spins down over X-ray quiescent periods \citep{1997ApJS..113..367B}.
This raises the question if the large spin-down of \sxp\ was observed by chance and if the source continued to spin down with such a high $\dot{P}_{\rm s}$.

An outburst of \sxp\ in 2012 was found in the optical with OGLE and in X-rays with the \swift\ satellite.
In this paper, we present our analysis of an \xmm\ target-of-opportunity observation of \sxp\ in Oct 2012, 
which allows to  measure the long-term evolution of the spin period of the NS on a longer time scale of $\sim$900 days since the first \xmm\ measurement.
All uncertainties in this paper are given for 1$\sigma$ confidence.

\begin{table*}
\caption[]{Spectral fit results.}
\centering
\begin{tabular}{lccccccccc}
\hline\hline\noalign{\smallskip}
\multicolumn{1}{l}{Model\tablefootmark{a}} &
\multicolumn{1}{c}{N$_{\rm H, smc}$} &
\multicolumn{1}{c}{$\Gamma$} &
\multicolumn{1}{c}{$kT$} &
\multicolumn{1}{c}{$N$\tablefootmark{b}} &
\multicolumn{1}{c}{$EW_{\rm Fe}$} &
\multicolumn{1}{c}{$F$\tablefootmark{c}} &
\multicolumn{1}{c}{$L$\tablefootmark{d}} &
\multicolumn{1}{c}{$\chi^2_{\nu}$} &
\multicolumn{1}{c}{dof} \\
\multicolumn{1}{c}{} &
\multicolumn{1}{c}{[\oexpo{21}cm$^{-2}$]} &
\multicolumn{1}{c}{} &
\multicolumn{1}{c}{[eV]} &
\multicolumn{1}{c}{[km]} &
\multicolumn{1}{c}{[eV]} &
\multicolumn{1}{c}{[erg cm$^{-2}$ s$^{-1}$]} &
\multicolumn{1}{c}{[erg s$^{-1}$]} &
\multicolumn{1}{c}{} &
\multicolumn{1}{c}{} \\
\noalign{\smallskip}\hline\noalign{\smallskip}
  PL               & 1.54$\pm$0.09  & 0.774$\pm$0.009 & --           & --                                       & --           &(6.8$\pm$0.7)\expo{-12} & 3.0\expo{36} & 1.01&1199  \\
  PL+BB            & 1.99$\pm$0.18  & 0.723$\pm$0.014 & 237$\pm$22   & 11.2$^{+2.7}_{-2.1}$ km                     & --           &(6.9$\pm$0.2)\expo{-12} & 3.1\expo{36} & 0.99&1197  \\
  PL+DiskBB        & 2.32$\pm$0.24  & 0.728$\pm$0.014 & 308$\pm$42   & 6.2$^{+1.4}_{-1.7}$ km                      & --           &(6.9$\pm$0.2)\expo{-12} & 3.1\expo{36} & 0.99&1197  \\
  PL+DiskBB+Fe     & 2.30$\pm$0.25  & 0.733$\pm$0.015 & 307$\pm$41   & 6.0$^{+2.3}_{-1.6}$ km                      & 40$\pm$13    &(6.9$\pm$0.2)\expo{-12} & 3.1\expo{36} & 0.98&1196  \\
  PL+APEC          & 1.78$\pm$0.11  & 0.729$\pm$0.011 & 1059$\pm$113 & $5.1^{+1.2}_{-0.9} \times 10^{57}$ cm$^{-3}$  & --           &(6.9$\pm$1.0)\expo{-12} & 3.1\expo{36} & 0.98&1197  \\
\noalign{\smallskip}\hline
\end{tabular}
\tablefoot{
All uncertainties are given for $\Delta\chi^2=1$ corresponding to a 1$\sigma$ confidence for one degree of freedom.
\tablefoottext{a}{For model definition see text.}
\tablefoottext{b}{Normalisation of the thermal model component: Radius (BB), inner disc radius for inclination of $\Theta=0$ (DiskBB), and emission measure (APEC).}
\tablefoottext{c}{Observed Flux in the (0.2--10.0) keV band.}
\tablefoottext{d}{Unabsorbed luminosity in the (0.2--10.0) keV band. A source distance of 60 kpc is assumed.}
}
\label{tab:spectra}
\end{table*}

\section{Observations and data reduction}
\label{sec:observations}

\begin{figure}
  \resizebox{\hsize}{!}{\includegraphics[angle=-90,clip=]{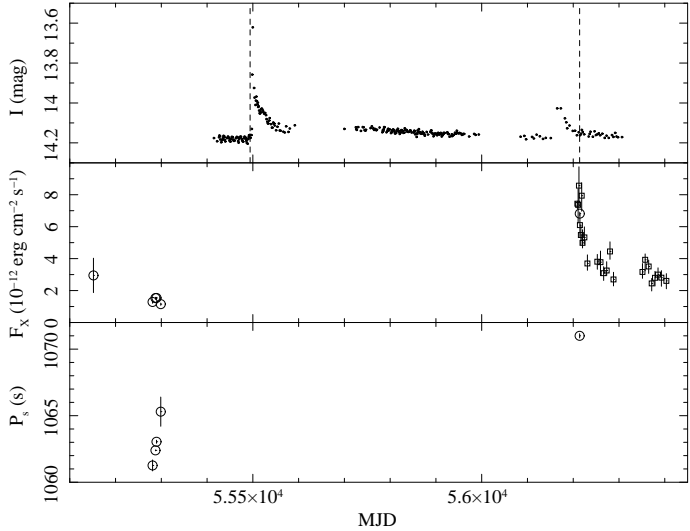}}
  \caption{
    {\it Upper panel:} OGLE-IV $I$-band light curve. Dashed lines indicate the time of optical spectroscopy observations.
    {\it Middle panel:} X-ray fluxes in the (0.2-10.0)~keV band from \swift\ (open squares) and \xmm\ (open circles) including the 2009 slew-survey data and the 2010 measurements.
    {\it Lower panel:} NS spin period as measured with \xmm.
    }
  \label{fig:lc}
\end{figure}

On 2012 Oct 9, \sxp\ was found in X-ray outburst with the \swift\ satellite. The evolution of the X-ray flux was followed in a monitoring campaign since then (TargetID: 32580).
The XRT spectra were reduced from the cleaned level-3 event files within a circle placed on the source
with  the {\tt ftool}\footnote{http://heasarc.nasa.gov/ftools/} {\tt xselect}.
Background spectra were created from a nearby point-source-free circular extraction region.
The ancillary response files were calculated with {\tt xrtmkarf}.

The detection of \sxp\ in X-ray bright state allowed us to ask for an \xmm\  \citep{2001A&A...365L...1J} follow-up observation (ObsID: 0700381801) that was performed on 2012 Oct 14.
\sxp\ was observed on-axis with all EPIC instruments in full-frame mode using the thin filter for EPIC-pn \citep{2001A&A...365L..18S} and the medium filter for both EPIC-MOS \citep{2001A&A...365L..27T}.
We processed the data using the \xmm\ SAS 12.0.1\footnote{Science Analysis Software (SAS), http://xmm.esac.esa.int/sas/}.
Because the flaring background was at moderate level (below 8 and 3 counts\,s$^{-1}$\,arcmin$^{-2}$ for pn and MOS, respectively), no temporal screening of the data was necessary.
This yields an exposure of 26.6, 31.1, and 31.2 ks for EPIC-pn, -MOS1, and -MOS2, respectively.
Energy spectra and time series of the source and the background were extracted, 
using the same selection regions as described in \citet[][see their Fig.~1, region A and B]{2012A&A...537L...1H} 
such that emission of the SNR will be subtracted.
We used single- and double-pixel events from EPIC-pn and single- to quadruple-pixel events from EPIC-MOS.
For energy spectra we rejected events with {\tt FLAG $\neq$ 0}.
This resulted in 24\,810, 8009, and 8908 net counts in the (0.2$-$12.0) keV band for the three instruments.
We binned the EPIC spectra to have a signal-to-noise ratio of $\geq 5$ for each bin. 
Spectra from the Reflection Grating Spectrometer \citep[RGS, ][]{2001A&A...365L...7D} with an exposure of 31.6 ks were extracted with {\tt rgsproc}.
We used events within a cross-dispersion point-spread-function width of 90\%, which excludes roughly half of the area covered by the SNR.
The background events were selected for a point-spread-function width $>$99\%, to exclude emission from the NS and the SNR.
The resulting spectra contain 277 (RGS1) and 380 (RGS2) net counts in the (0.35$-$2.0) keV band.
The RGS spectra were binned with 25 net cts bin$^{-1}$.
Although these spectra have low statistics, they provide the first grating X-ray data of \sxp.

We used the Southern African Large Telescope  \citep[SALT, ][]{2006SPIE.6267E..32B,2006SPIE.6269E...8B,2006MNRAS.372..151O} to collect long slit spectra of \sxp\ simultaneous with the \xmm\ observation. 
As part of the SALT program 2012-1-RSA\_UKSC-003 (PI: Schurch) aimed at studying outbursts from X-ray binary systems, we obtained spectra of \sxp\ during the evening of 2012 Oct 13th and morning of the 14th. 
These data were collected using the Robert Stobie Spectrograph \citep[RSS, ][]{2003SPIE.4841.1463B,2003SPIE.4841.1634K}.  
We used the PG2300 grating in two position angles (30.500\degr, and 48.875\degr) to achieve coverage from (3835$-$4924) \AA\ in the blue and (6082$-$6925) \AA\ in the red, respectively.  
To achieve the highest resolution we used a 0.6\arcsec\ wide slit. Observations were prebinned by a factor 2 producing a spectral resolution of 0.33 \AA\ per binned pixel.  
Two 400 s observations were taken in the blue on MJD~56213.99 followed by a 180 s exposure of a ThAr arc lamp.  
A single 180 s exposure was taken in the red on MJD~56214.01 followed by a 90 s exposure of an Ar arc lamp.  
All observations were flat fielded using Quartz Tungsten Halogen (QTH) flats taken immediately after the blue and before the red observations.  
The image quality during the observations was approaching 3\arcsec.
An hour after the observations a spectrophotometric standard star (LTT1020) was observed for both the blue and red setups.
The SALT data was first processed using the SALT pipeline, {\tt PySALT} \citep{2010SPIE.7737E..54C}. 
This performs overscan, gain, cross-talk corrections, and mosaicing of the three 2048$\times$4096 CCDs.  
Subsequent flatfielding, background subtraction, wavelength calibration and extraction of 1D spectra were performed in {\tt IRAF} v2.16 \citep{1993ASPC...52..173T}.  
The final spectra were first redshift corrected by 150 km s$^{-1}$ and normalised (after being averaged in the case of the two blue 400 s exposures).

An optical monitoring of \sxp\ is provided by the Optical Gravitational Lensing Experiment \citep[OGLE, ][]{2008AcA....58...69U} 
that covers the source regularly with $I$-band photometry since the beginning of phase IV (2010 Aug 6).
The data is available in the X-Ray variables OGLE Monitoring \citep[XROM, ][]{2008AcA....58..187U} system and presented in the upper panel of Fig.~\ref{fig:lc}.

\section{Analyses and Results}
\label{sec:analyses}

\subsection{X-ray energy spectrum}
\label{sec:analyses:xspec}

\begin{figure}
  \resizebox{\hsize}{!}{\includegraphics[angle=-90,clip=]{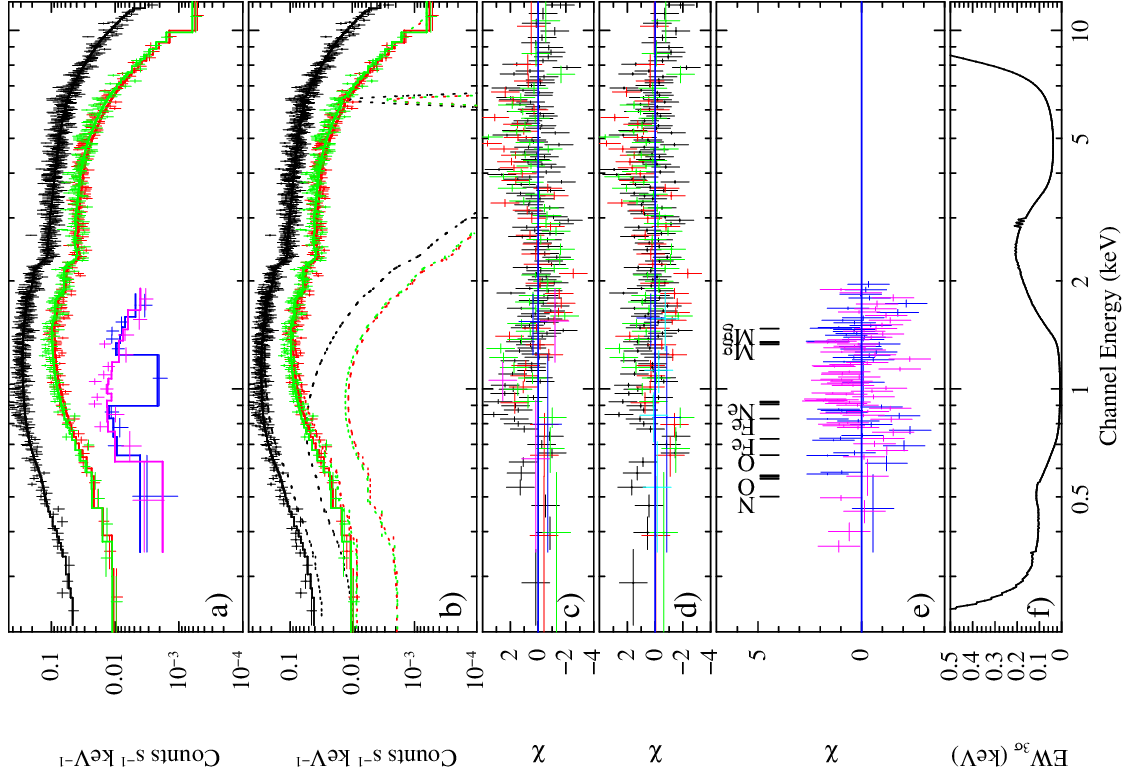}}

  \caption{
    {\it a)} Energy spectrum of \sxp\ as observed on 2012 Oct 14 with EPIC-pn (black), -MOS1 (red), -MOS2 (green), RGS1 (blue), and RGS2 (magenta)
             together with the best-fit model (solid line) of an absorbed power law.
    {\it b)} Same as {\it a)}, but fitted with additional contribution of a multi-temperature disc black-body model and Fe K$_{\alpha}$ emission line, shown by dashed lines.
             The RGS data is omitted for clarity.
    {\it c)} The residuals in units of $\sigma$ for the model from {\it a)} with a higher binning by a factor of 5.
    {\it d)} Same as {\it c)}, but for the model from {\it b)}.
    {\it e)} The residuals for the model of {\it a)}, binned with only 5 net cts bin$^{-1}$.
             Vertical lines mark the energies of prominent emission lines.
    {\it f)} 3$\sigma$ upper limits for the equivalent width ($EW$) of a Gaussian absorption line.
    }
  \label{fig:spectrum}
\end{figure}

Spectral analysis was performed with {\tt xspec} \citep{1996ASPC..101...17A} version 12.7.0u.
We used the same model, as described in \citet{2012A&A...537L...1H}, i.e. an absorbed power law 
where we assume Galactic photoelectric absorption with a column density of N$_{\rm H, gal}=6 \times 10^{20}$ cm$^{-2}$ with solar abundances according to \citet{2000ApJ...542..914W},
and a free column density, N$_{\rm H, smc}$, with abundances set to 0.2 solar accounting for absorption by the interstellar medium of the SMC or within the BeXRB system.
The normalisations between the individual instruments were found to be consistent, and we fitted the same model to all five spectra simultaneously.
The spectrum and best-fit model are presented in Fig.~\ref{fig:spectrum}a and the best-fit parameters are listed in Table~\ref{tab:spectra}.
Formally, the power-law model describes the data well with a reduced $\chi^2_{\nu}$ of 1.01. 
However, in other BeXRBs, a soft excess \citep{2004ApJ...614..881H} and iron fluorescent emission are believed to contribute to the X-ray spectrum.

A possible Fe K$_{\alpha}$ emission line broadened below the instrumental resolution was investigated by fitting an
additional Gaussian emission-line profile with central energy of $E_{\rm c}=6.4$ keV, line width of $\sigma=0$ and free normalisation.
We obtain a 3.1$\sigma$ evidence for a line with normalisation of  $N=(3.1\pm1.0)\times 10^{-6}$ photons cm$^{-2}$ s$^{-1}$, corresponding to an equivalent width of $EW = (-40\pm13)$ eV.
Analogously, we derive for ionised \ion{Fe}{xxv} ($E_{\rm c}=6.7$ keV) a 3$\sigma$ upper limit of $N<2.9\times 10^{-6}$ photons cm$^{-2}$ s$^{-1}$ 
and $N<3.1\times 10^{-6}$ photons cm$^{-2}$ s$^{-1}$ for \ion{Fe}{xxvi} ($E_{\rm c}=7.0$ keV).

To investigate a possible soft excess and to demonstrate systematic uncertainties of the power-law and absorption parameters we fitted additional thermal components.
Adding a black-body (BB) emission component to the model results in a slight improvement of the fit with an f-test probability of $2.3\times 10^{-7}$.
This formally proves the significance of this component, however see \citet{2002ApJ...571..545P} for limitations of the f-test.
We further fitted a multi-temperature accretion disc model (DiskBB) and a collisionally ionised plasma model with 0.2 solar abundances (APEC).  The results are presented in Table~\ref{tab:spectra}.
The possible contribution of the disc black-body model and Fe K$_{\alpha}$ line are demonstrated in Fig.~\ref{fig:spectrum}b, where we omit the RGS spectra for clarity. 
The residuals of this model, rebinned by a factor of 5, are compared with the residuals of the simple power-law model in Fig.~\ref{fig:spectrum}d and c. 
We find that e.g. in the case of a disc black-body a soft excess can only contribute with a luminosity of  
$7.5_{-1.1}^{+0.8} \times 10^{34}$ erg s$^{-1}$, i.e. $2.5_{-0.4}^{+0.3} \%$ of the unabsorbed luminosity in the (0.2$-$10.0) keV band.

To check for spectral features in the RGS grating spectra, we plot the residuals of the power-law fit with a binning of only 5 net cts bin$^{-1}$.
We find clear indications of emission lines of the helium-like triplets of 
\ion{O}{vii}  (0.561$-$0.574 keV), 
\ion{Ne}{ix}  (0.905$-$0.922 keV), and 
\ion{Mg}{xi}  (1.33$-$1.35 keV)
as well as hydrogen-like 
\ion{N}{vii}  (0.500 keV),
\ion{O}{viii} (0.654 keV), and
\ion{Mg}{xii} (1.47 keV)
and neon-like \ion{Fe}{xvii} (0.725, 0.826 keV), as marked in Fig.~\ref{fig:spectrum}e.
Due to the low statistics in the spectra, we can only roughly constrain the fluxes of the most convincing lines to 
$1.6^{+0.9}_{-0.8} \times 10^{-5}$ photons cm$^{-2}$ s$^{-1}$ (\ion{O}{vii}),
$4.9_{-2.5}^{+3.4} \times 10^{-6}$ photons cm$^{-2}$ s$^{-1}$ (\ion{O}{viii}), and 
$1.3_{-0.4}^{+0.5} \times 10^{-5}$ photons cm$^{-2}$ s$^{-1}$ (\ion{Ne}{ix}).
Here we used the unbinned data and C statistics \citep{1979ApJ...228..939C}.

The peculiar properties of \sxp\ were suggested to be caused by a strong magnetic field \citep[$B \sim 10^{14}$ G, ][]{2012ApJ...757..171F},
possibly leading to a proton cyclotron feature in the observed energy band.
Adding a Gaussian absorption line to the power-law model, we find no significant contribution of this feature. 
Stepping the absorption line through the complete spectrum, we estimate 3$\sigma$ upper limits for $EW$ as shown in Fig.~\ref{fig:spectrum}f, where we assume a line width of $\sigma=0.2 E_{\rm c}$.

\begin{figure}
  \resizebox{\hsize}{!}{\includegraphics[angle=-90,clip=]{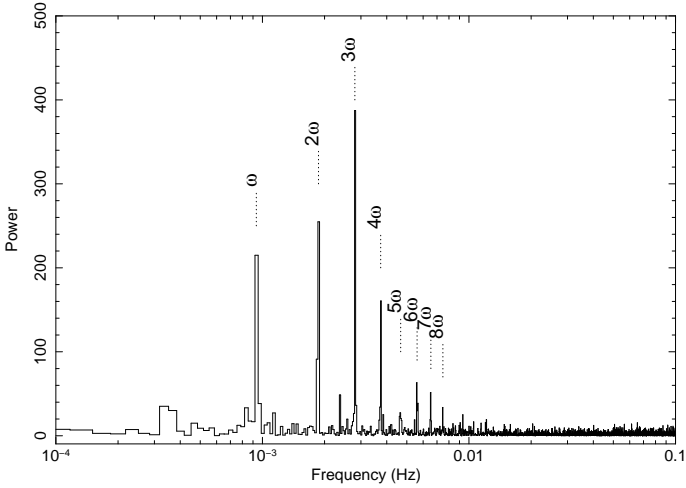}}

  \caption{
    Power-density spectrum of \sxp. The best-fit frequency $\omega=934$ $\mu$Hz and harmonics are marked by vertical lines.
  }
  \label{fig:fft}
\end{figure}

\begin{figure}
  \resizebox{\hsize}{!}{\includegraphics[angle=0,clip=]{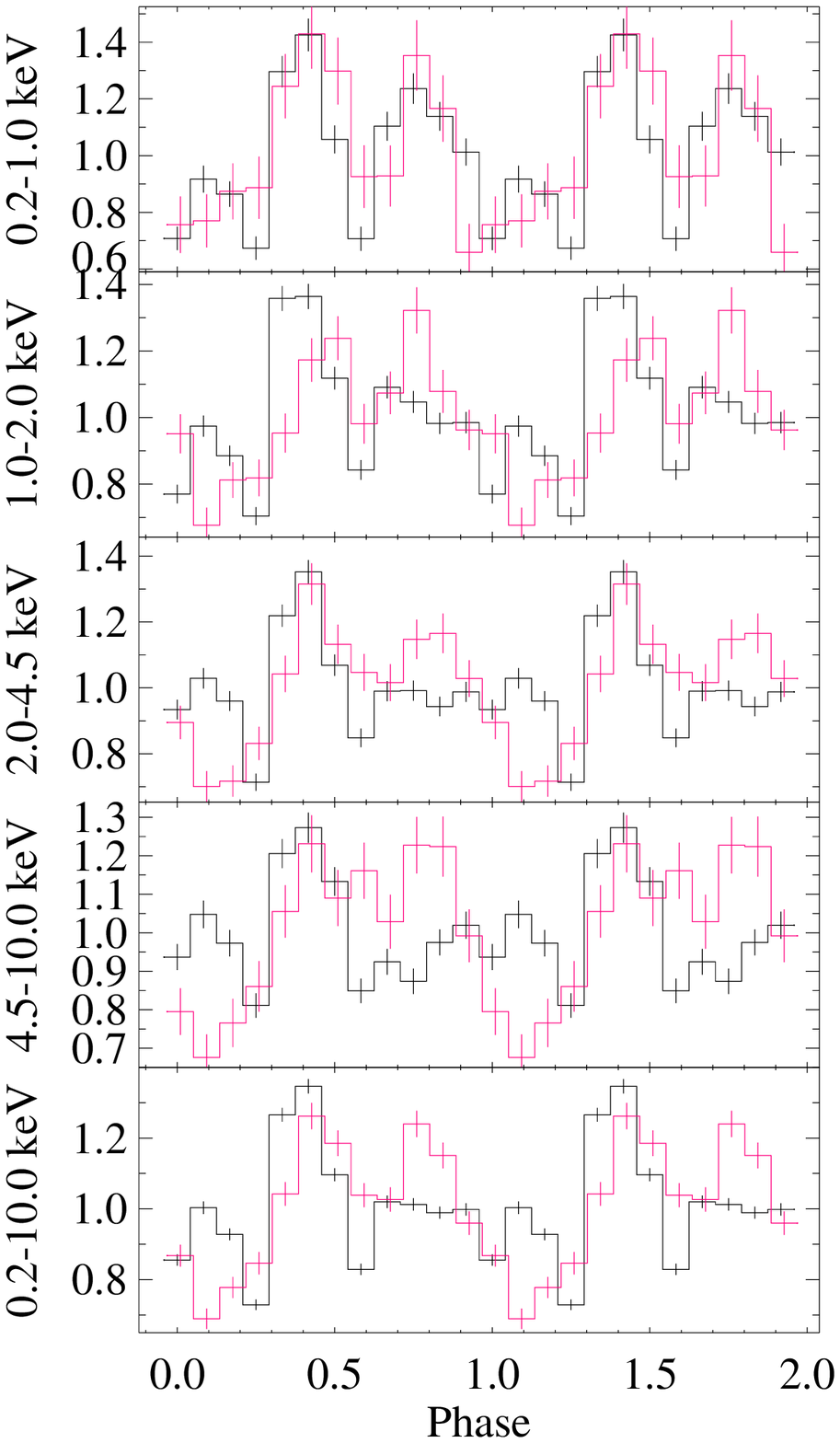}}
  \resizebox{\hsize}{!}{\includegraphics[angle=0,clip=]{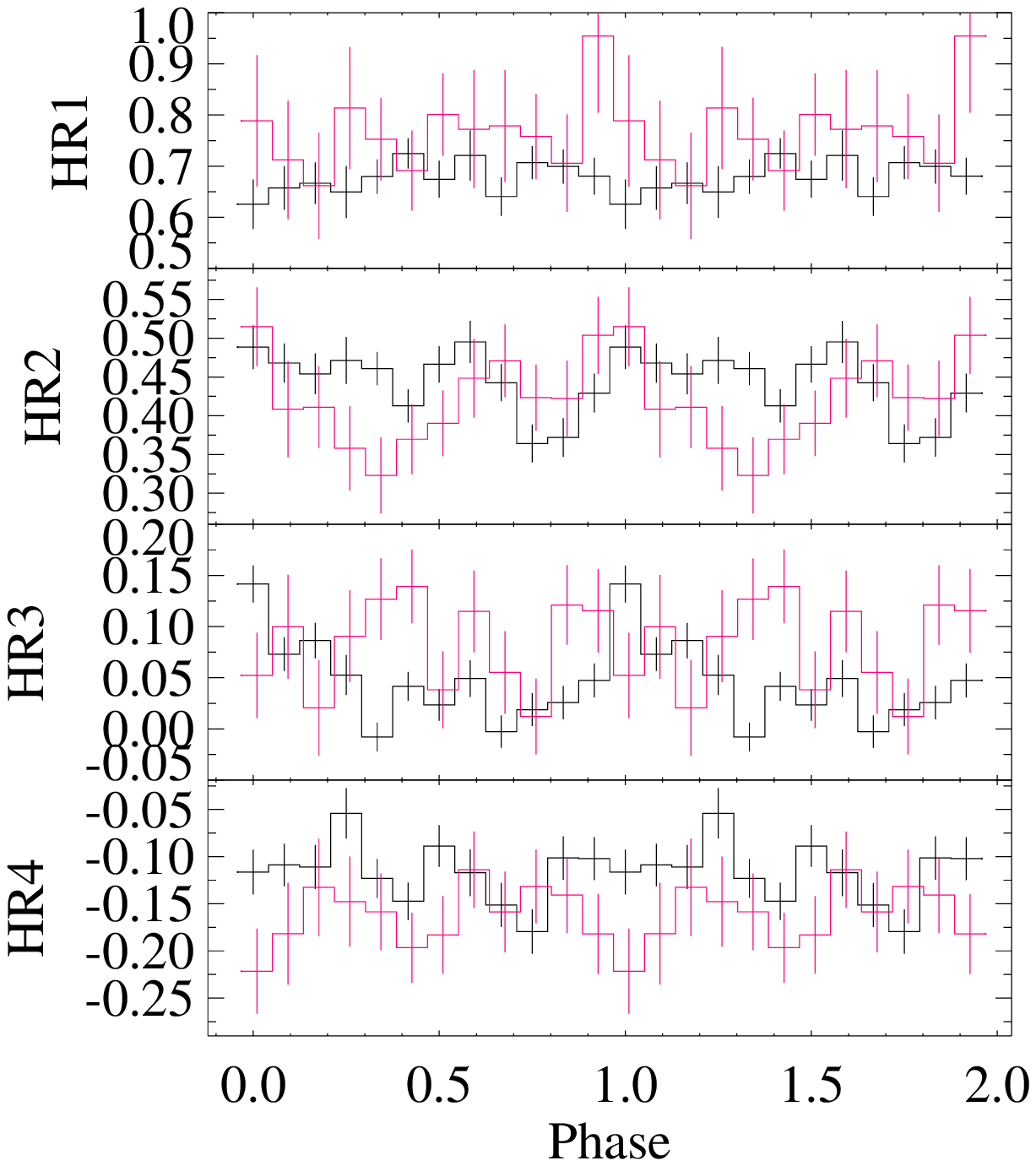}}
  \caption{
    {\it Upper figure:}  Pulse profile of the time series, merged from all EPIC instruments in different energy bands for the 2012 observation (black) and an observation in 2010 (red).
                 The individual light curves are background subtracted and normalised to the average count rate 
                 ({\it from top to bottom:} 0.19, 0.40, 0.44, 0.35, 1.37 cts s$^{-1}$ (2012)  
                  and                       0.020, 0.044, 0.052, 0.037, 0.15 cts s$^{-1}$ (2010)).
    {\it Lower figure:} Hardness ratios as function of pulse phase. 
   }
  \label{fig:pp}
\end{figure}

\subsection{X-ray Pulsations}

A fast Fourier transformation of the merged EPIC time series (42\,231 cts including background) reveals the pulsation of the NS and various harmonics as presented in Fig.~\ref{fig:fft}.
Using a Bayesian detection method as described by \citet{1996ApJ...473.1059G} and \citet{2008A&A...489..327H}, the most probable spin period during the recent \xmm\ observation is derived to $P_{\rm s} =(1071.01\pm0.16)$ s.

The background-subtracted normalised folded light curves are shown in Fig.~\ref{fig:pp} for the total (0.2$-$10.0) keV band 
and the standard sub-bands (0.2$-$0.5) keV, (0.5$-$1.0) keV, (1.0$-$2.0) keV, (2.0$-$4.5) keV, and (4.5$-$10.0) keV,
where we merged the first two bands to increase the statistics.
The HRs are defined by HR$_{i}$ = (R$_{i+1}$ $-$ R$_{i}$)/(R$_{i+1}$ + R$_{i}$) with R$_{i}$ denoting the background-subtracted 
count rate in the \xmm\ standard energy band $i$ (with $i$ from 1 to 4). 
In Fig.~\ref{fig:pp}, we also show the pulse profile measured in 2010 (ObsID 0602520201), folded with 1062.4 s.
The phase offsets are set to have a common maximum in the total-band light curves.

\subsection{Long-term X-ray light curve}
\label{sec:X-rayLC}

The position of \sxp\ was covered with the {\it Einstein} satellite, but no source is reported in the catalogue of \citet[][]{1992ApJS...78..391W}.
Also the ASCA catalogue of \citet{2003PASJ...55..161Y} does not contain \sxp,
which was in the FoV of an observation performed on 1999 May 28 to 29.

We note that both catalogues list a nearby source 
(No 67 of \citet[][]{1992ApJS...78..391W} and No 105 of \citet{2003PASJ...55..161Y}).
The angular separation of $\sim$4.4\arcmin\ is too large for a correlation with \sxp\ with respect to the angular resolution of both satellites.
In the recent \xmm\ observation, as well as in the 2010 observations, we do not find a bright source at this position.
Hence, the nearby source might be an X-ray transient. In the case of another BeXRB, the closest possible counterparts are 
2dFS\,3857 (B0-5V star, 60\arcsec angular separation), and 
2MASS J01281201-7330235 (B3-5III star showing near-infrared variability, 80\arcsec).

Three pointed ROSAT observations of SMC\,X-1 covered the position of \sxp\ between 1991 and 1993 (MJD 48536.2,  48895.7, and 49141.0). 
For these observations, no detection at the position of \sxp\ is reported in the catalogue of \citet{2000A&AS..142...41H}.
Assuming 10 counts as a typical detection limit and the same spectral shape as measured in the recent \xmm\ observation, 
we derive upper limits of 1.4, 2.7, and 2.1 $\times 10^{-13}$ erg cm$^{-2}$ s$^{-1}$, respectively,
which are the lowest limits for the X-ray flux reported so far.

As noted by \citet{2012MNRAS.420L..13H}, the very first X-ray detection of \sxp\ is listed in the \xmm\ slew-survey catalogue \citep[][ Release 1.5]{2008A&A...480..611S}.
The source (XMMSL1\,J012746.2-733304) was detected on 2009 Nov 16 with ($4.2\pm1.5$) cts, which according to the spectrum from above, results in a flux of $(3.0\pm1.1)\times 10^{-12}$ erg cm$^{-2}$ s$^{-1}$.

The recent \swift/XRT monitoring allows to follow the evolution of the X-ray outburst in 2012.
Since the X-ray spectrum is measured well with \xmm\ and we find no strong variations of the spectral shape compared to the 2010 observations,
we fixed the spectral shape to the PL+DiskBB+Fe model and calculated fluxes for each \swift\ spectrum, using C statistics.
The derived fluxes are listed in Table~\ref{tab:swift-obs}.
The X-ray light curve is presented in Fig.~\ref{fig:lc} and compared to the optical light curve from OGLE and the evolution of the spin period of the NS.

\begin{table}
 \caption{\swift/XRT observations of \sxp.}
 \begin{center}
   \begin{tabular}{lccccrrrr}
     \hline\hline\noalign{\smallskip}
     \multicolumn{1}{c}{MJD\tablefootmark{a}} &
     \multicolumn{1}{c}{Net Exp} &
     \multicolumn{1}{c}{Rate} &
     \multicolumn{1}{c}{Flux\tablefootmark{b}} \\
     \multicolumn{1}{c}{} &
     \multicolumn{1}{c}{[s]} &
     \multicolumn{1}{c}{[$10^{-2}$ s$^{-1}$]} &
     \multicolumn{1}{c}{[$10^{-12}$ erg cm$^{-2}$ s$^{-1}$]} \\
     \noalign{\smallskip}\hline\noalign{\smallskip}
56209.01  &  15558   & 6.49$\pm$0.20    &$7.44_{-0.31}^{+0.17}$ \\ \noalign{\smallskip}      
56210.75  &  976     & 6.25$\pm$0.80    &$7.37_{-1.13}^{+0.99}$ \\ \noalign{\smallskip}      
56212.48  &  1019    & 7.36$\pm$0.85    &$8.57_{-1.06}^{+1.18}$ \\ \noalign{\smallskip}      
56214.62  &  1151    & 5.30$\pm$0.68    &$6.10_{-0.72}^{+0.66}$ \\ \noalign{\smallskip}      
56216.69  &  2078    & 4.72$\pm$0.48    &$5.48_{-0.60}^{+0.62}$ \\ \noalign{\smallskip}      
56218.56  &  1269    & 6.46$\pm$0.71    &$7.94_{-0.97}^{+0.81}$ \\ \noalign{\smallskip}      
56220.37  &  1748    & 4.06$\pm$0.48    &$4.99_{-0.32}^{+0.82}$ \\ \noalign{\smallskip}      
56224.44  &  1823    & 4.44$\pm$0.49    &$5.33_{-0.42}^{+0.66}$ \\ \noalign{\smallskip}      
56231.03  &  1723    & 2.96$\pm$0.41    &$3.70_{-0.43}^{+0.54}$ \\ \noalign{\smallskip}      
56252.70  & 1776     & 3.32$\pm$0.43    &$3.81_{-0.47}^{+0.46}$ \\ \noalign{\smallskip}      
56259.58  & 729      & 3.15$\pm$0.66    &$3.78_{-0.85}^{+0.70}$ \\ \noalign{\smallskip}      
56266.45  & 1895     & 2.22$\pm$0.34    &$3.09_{-0.45}^{+0.41}$ \\ \noalign{\smallskip}      
56273.06  & 2000     & 2.70$\pm$0.37    &$3.26_{-0.33}^{+0.55}$ \\ \noalign{\smallskip}      
56280.33  & 2188     & 3.93$\pm$0.42    &$4.45_{-0.49}^{+0.60}$ \\ \noalign{\smallskip}      
56288.01  & 1686     & 2.14$\pm$0.36    &$2.70_{-0.41}^{+0.45}$ \\ \noalign{\smallskip}      
56351.36  & 1940     & 2.83$\pm$0.38    &$3.17_{-0.41}^{+0.52}$ \\ \noalign{\smallskip}      
56357.17  & 1438     & 3.41$\pm$0.49    &$3.93_{-0.69}^{+0.37}$ \\ \noalign{\smallskip}      
56364.90  & 1878     & 3.14$\pm$0.41    &$3.51_{-0.43}^{+0.40}$ \\ \noalign{\smallskip}      
56371.51  & 2123     & 2.17$\pm$0.32    &$2.45_{-0.47}^{+0.31}$ \\ \noalign{\smallskip}      
56378.71  & 1950     & 2.36$\pm$0.35    &$2.80_{-0.51}^{+0.41}$ \\ \noalign{\smallskip}      
56385.40  & 1945     & 2.78$\pm$0.38    &$3.00_{-0.32}^{+0.42}$ \\ \noalign{\smallskip}      
56392.73  & 1201     & 2.66$\pm$0.47    &$2.81_{-0.54}^{+0.44}$ \\ \noalign{\smallskip}      
56403.51  & 1456     & 2.40$\pm$0.41    &$2.61_{-0.50}^{+0.50}$ \\ \noalign{\smallskip}      
      \noalign{\smallskip}\hline\noalign{\smallskip}
    \end{tabular}
 \end{center}
  \tablefoot{
  \tablefoottext{a}{Modified Julian date of the beginning of the observation.}
  \tablefoottext{b}{Flux in the (0.2$-$10.0) keV band.}
  }
 \label{tab:swift-obs}
\end{table}

\subsection{Optical spectroscopy}

\begin{figure*}
  \resizebox{\hsize}{!}{\includegraphics[angle=0,clip=]{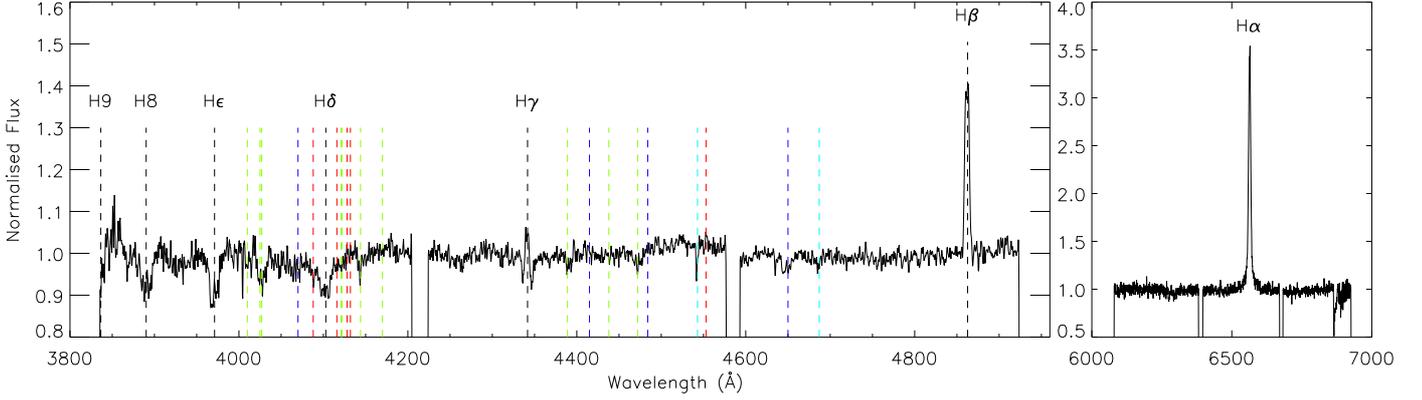}}

  \caption{SALT blue and red (left and right respectively) smoothed spectra of \sxp. 
           Clearly visible are the chip gaps between the three CCDs.  
           Dotted lines indicate: Balmer lines (black), \ion{He}{i} (green), \ion{He}{ii} (light blue), Silicon (red) and other metal lines (dark blue).
          }
  \label{fig:opt_spectrum}
\end{figure*}

\begin{figure*}
  \resizebox{\hsize}{!}{
\includegraphics[height=0.3\textheight,angle=0,clip=]{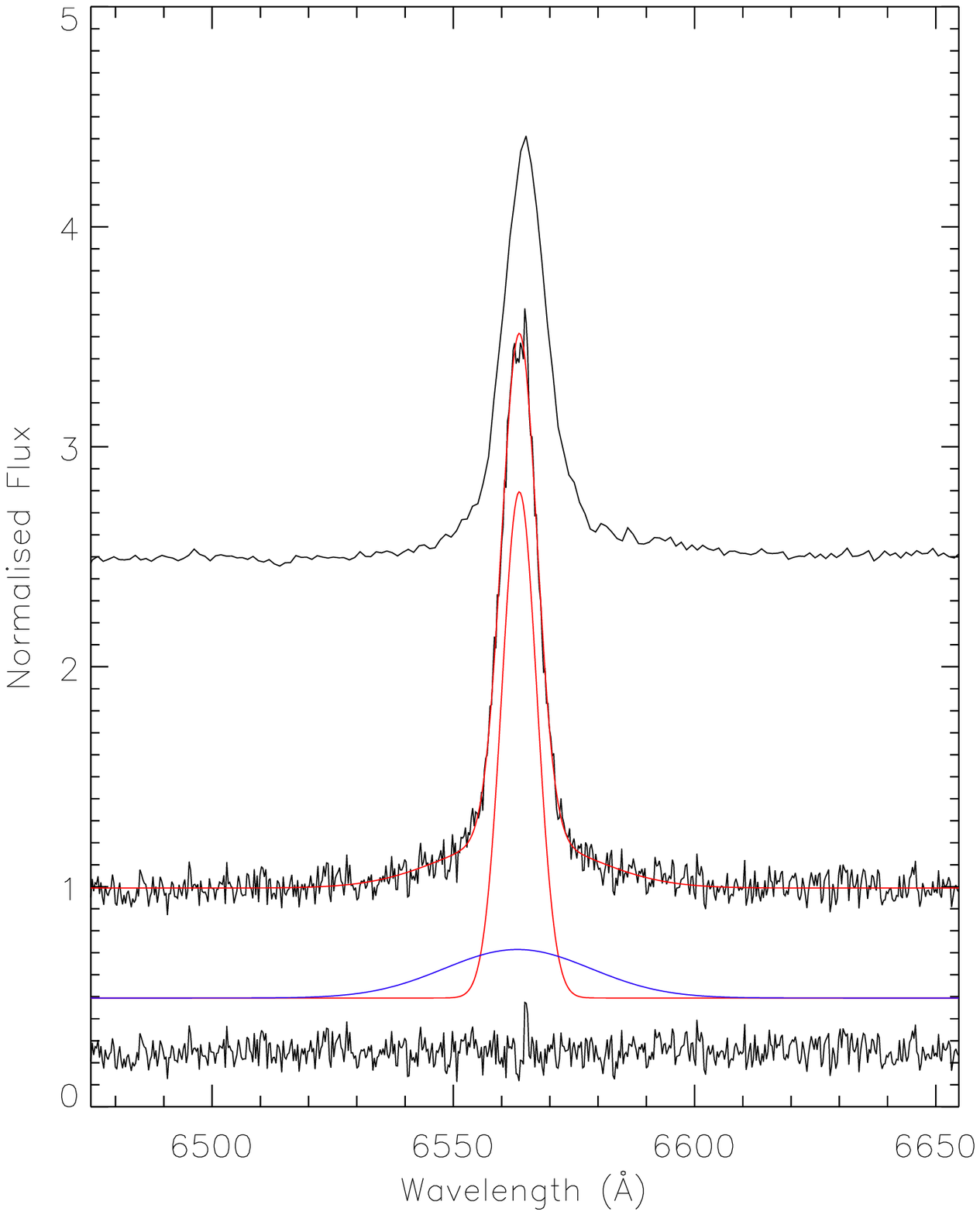}
\includegraphics[height=0.3\textheight,angle=0,clip=]{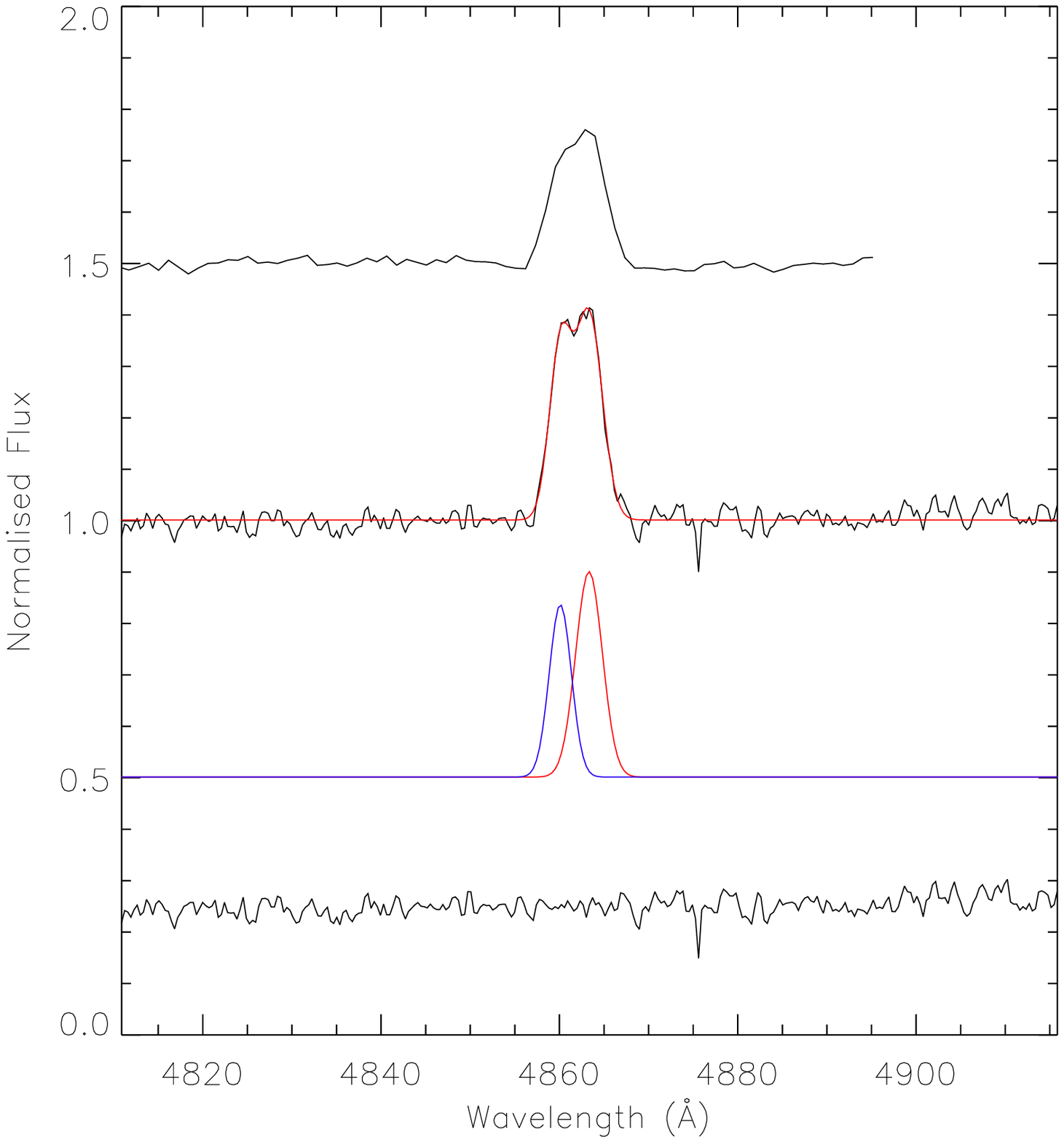}
\includegraphics[height=0.3\textheight,angle=0,clip=]{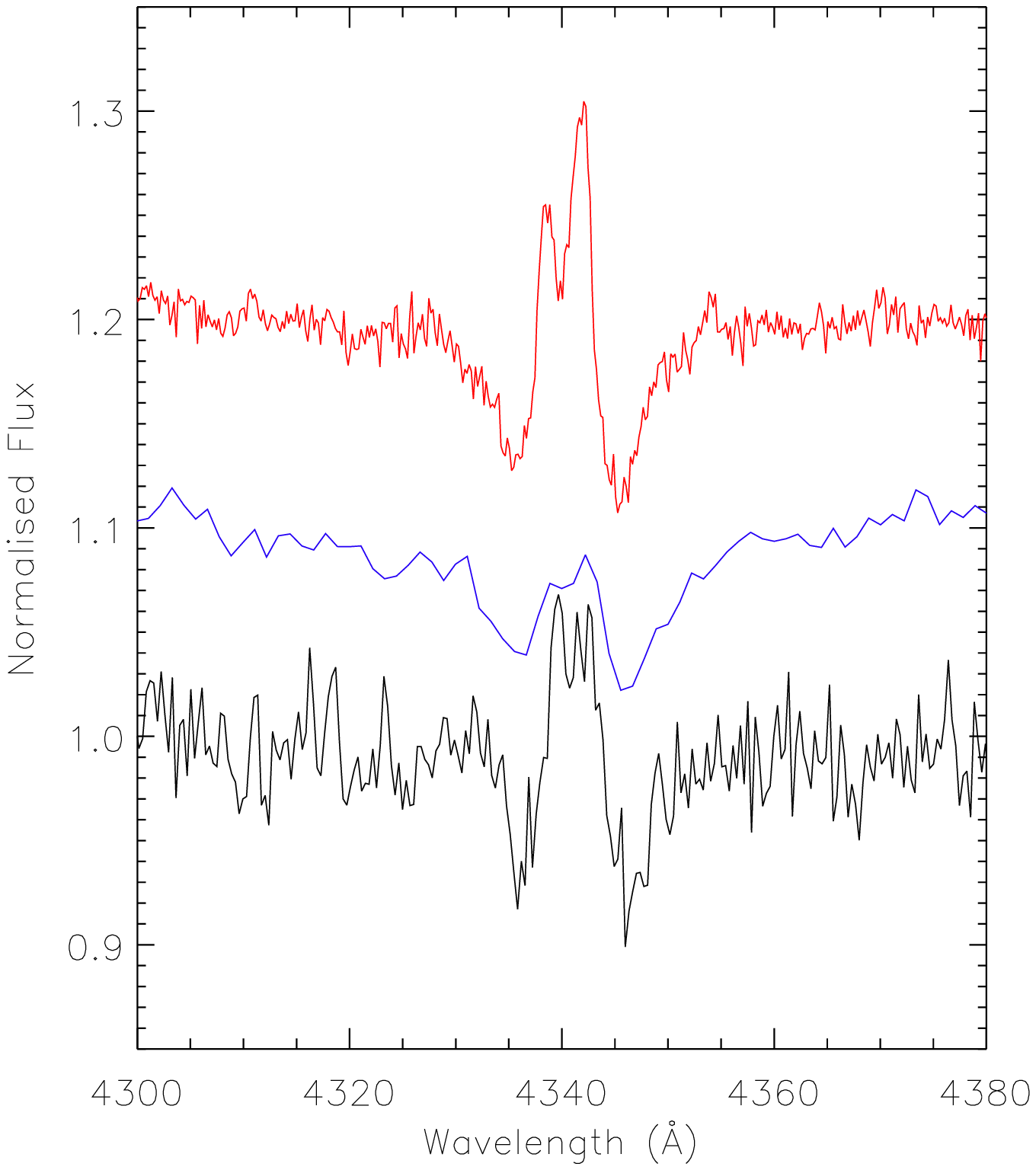}
}
  \caption{{\it Left:} H$\alpha$ as seen with SALT (normalised) and according modelling with two Gaussians and continuum. The lowest line gives the residuals. The 2dF spectrum is shown on the top for comparison.
           {\it Middle:} Same as before, but for H$\beta$.
           {\it Right:} H$\gamma$ as seen with VLT FLAMES (top), 2dF (middle) and SALT (bottom).
          }
  \label{fig:opt_spectrum_zoom}
\end{figure*}

Fig.~\ref{fig:opt_spectrum} shows the fully reduced spectra after being smoothed using a boxcar average of 3 data points.
The H$\alpha$ and H$\beta$ line are shown in more detail in Fig.~\ref{fig:opt_spectrum_zoom}.
Neither line could be fitted well using one Gaussian alone. 
For H$\alpha$, broad wings are seen in addition. 
H$\beta$ clearly exhibits a double-peaked line profile that is also seen in the pre-subtraction images.
Fitting both lines with two Gaussians each and using the unsmoothed data, 
we derive equivalent widths of the H$\alpha$ and H$\beta$ of $EW_{\rm H \alpha}=(-26.65\pm0.09)$ \AA\ and $EW_{\rm H \beta}=(-2.40\pm0.29)$ \AA, respectively.
Re-analysing previous 2dF spectra \citep{2004MNRAS.353..601E,2012MNRAS.420L..13H} with the same method, 
results in $EW_{\rm H \alpha}=(-22.02\pm0.054)$ \AA\ and  $EW_{\rm H \beta}=(-1.58\pm0.10)$ \AA.

\section{Discussion}
\label{sec:discussion}

We analysed an \xmm\ observation of \sxp\ and find that the spin period of the NS further increased to 1071 s.
Since the last \xmm\ observation, performed on 2010 Apr 12, 915 days before the most recent observation, we obtain an average $\dot{P}_{\rm s} = (2.27\pm0.44)$ s yr$^{-1}$.
This is factor of 40 less than observed during the 18 day baseline in 2010.
By assuming, that this average $\dot{P}_{\rm s}$ is representative for the NS since birth over 25 kyr, one derives an initial $P_s$ of $\sim$20 s.
If one assumes that the NS is born with a $P_s$ of a few ms, a high initial magnetic field strength of $\sim$$10^{14}$ G \citep[compared to a few $10^{12}$ G as typically found for BeXRBs, e.g.][]{2012AIPC.1427...60P}
and an efficient propeller mechanism are needed \citep[e.g. ][]{2012ApJ...757..171F} to slow the pulsar down within a few kyr to be close to the equilibrium period now.

The more moderate long-term spin-down rate suggests that the source was observed during high spin down phase in 2010 by chance 
and that the high spin-down rate is not an intrinsic feature of this source.
This supports the hypothesis that \sxp\ is rotating close to its equilibrium period, with spin-up during periods of high accretion rates and spin-down at other times.
In this case, we would expect to observe spin-up of the NS occasionally as well, however there is no observational evidence so far.
The recent data do not allow to conclude on the detailed evolution
as periods of spin-up and -down may have alternated as seen in other BeXRBs.
Especially the current $\dot{P}_{\rm s}$ cannot be determined from our single observation and the NS might even have shown spin-up during the X-ray outburst monitored by \swift.
A denser X-ray coverage of the source is highly desirable to determine the evolution of $P_{\rm s}$ and $\dot{P}_{\rm s}$ on a longer time scale and to constrain individual models.

Comparing the pulse profiles of 2010 and 2012,
we find indications for a double peaked light curve in 2010, 
whereas in 2012 the light curve reveals only one clear peak at energies above 1 keV.

The shape of the X-ray spectrum is close to the 2010 measurement, but might contain a soft excess and an Fe K line.
If the soft excess is caused by black-body emission, the black-body radius is in agreement with the radius of an NS and might originate from the NS surface.
But also reprocessing of hard X-rays from the NS in nearby material like an accretion disc is possible, as we can only measure a lower limit of the inner disc radius, depending on the inclination ($R \sim \cos^{-1/2}{\Theta}$).
Estimating the inner disc radius according to \citet{2004ApJ...614..881H} yields $R=(L_{\rm X}/ 4\pi\sigma T^4)^{1/2}  = 8.5$ km.

The high-resolution X-ray spectra from the RGS show indications of emission lines.
The emission lines might be explained by the SNR around \sxp.
However, assuming a homogeneous surface brightness of the SNR and typical SMC abundances of $Z=0.2Z_{\sun}$,
the best-fit plasma model for the SNR of \citet{2012A&A...537L...1H} accounts only for a few percent of the observed \ion{O}{vii} line flux.
This points to an additional hot plasma component around the NS as it was e.g. observed in SXP\,1323 \citep{2005A&A...438..211H}.
Deeper X-ray observations of the source in an X-ray bright state are needed here,
but we note that the line energies of the helium-like lines point to resonance lines rather than forbidden lines
and that the emission of the observed He-like species is stronger than H-like like.

Proton cyclotron absorption lines \citep[e.g. ][]{2001ApJ...560..384Z} can be significant features in the X-ray spectra of NSs with high magnetic fields as observed with present-day observatories
and might appear in the X-ray spectrum of \sxp\ \citep[][]{2012ApJ...757..171F}.
Neither in the residuals of the simple power-law fit (Fig.~\ref{fig:spectrum} c)
nor by adding an absorption-line component to the model (Sec.~\ref{sec:analyses:xspec}) 
we can find a convincing contribution of an absorption line.
A weak absorption line with equivalent width down to $EW>10$ eV can be excluded only in the (0.84$-$1.17) keV band,
whereas a stronger line with $EW>200$ eV can be ruled out in the (0.28$-$7.5) keV band.
According to the relation

$$ B\simeq 1.59\times 10^{14}\: z_{\rm G}^{-1} \left(\frac{E_{\rm c}}{\rm keV}\right) \: {\rm G}, $$ 
where $z_{\rm G}$ is the gravitational redshift,
this corresponds to magnetic fields in the range of  $(1.3-1.9)\times 10^{14}$ G  and $(0.44-12)\times 10^{14}$ G, respectively%
, for $z_{\rm G}=1$.
However, we note that also no convincing detection of a proton cyclotron-absorption feature has been reported for magnetars up to now \citep{2008A&ARv..15..225M}.

The long term $I$-band light curve, as measured with OGLE-IV has been examined by \citet{2012ATel.4399....1S} and exhibits short term variability of $\sim 9$ days, 
that is likely an alias of sinusoidal modulations of $(0.9007\pm0.0005)$ d, caused by non-radial pulsations.
In addition, two strong outbursts are seen reaching maximum emission around MJD 55500 ($\Delta I \sim 0.6$ mag) and between MJD 56151 -- 56173 ($\Delta I \gtrsim 0.15$ mag).
\citet{2012ATel.4596....1S} suggested the time-difference between the outbursts to be the likely orbital period $P_{\rm o} = (656\pm2)$ d.
This is supported by the FRED-like shape \citep{2012MNRAS.423.3663B} of the optical outbursts.
If so, the X-ray emission in 2010 cannot be caused by a type-I X-ray outburst that can occur during periastron passage of the NS only.

This might be explained by a persistent X-ray state that is also favoured by the fact, that
\sxp\ was always detected above a flux of $\sim 10^{-12}$ erg cm$^{-2}$ s$^{-1}$ since its discovery.
The pulsar must have entered this state after the last ROSAT observation.
Comparing the deepest ROSAT upper limit of $F_{\rm X} \leq 1.4 \times 10^{-13}$ erg cm$^{-2}$ s$^{-1}$ with the average flux of the 2010 \xmm\ observations reveals an X-ray variability by a factor of 10 
and by comparing with the brightest \swift\ detection, one obtains a maximum X-ray variability by a factor of at least 60.
The X-ray flux is also moderately (some 10\%) variable on short time scales (days), as measured with \swift\ and previously with \cxo\ \citep[c.f. Fig. 2 of][]{2013IAUS..291..459O}.

The X-ray outburst around MJD 56200 may correlate with the second optical outburst.
Because our \swift\ observations did not catch the initial stage of X-ray outburst we cannot judge about the outburst duration and maximum emission.
The first \swift\ observation was performed 58 days after the first measured $I$-band increase.
Regarding the long orbital period, this time interval might still be consistent with a type-I X-ray outburst.
In this scenario, the NS accretes matter during periastron passage, leading to X-ray emission, while tidal forces affect the circumstellar disc.
A sudden but local perturbation in the disc geometry, such as e.g. warping a part of the disc or the formation of a 
quasi-steady accretion disc around the NS, might cause the sharp increase in total optical luminosity.  
Crude estimates show that the observed increase in $I$-band magnitude can be achieved if the projected disc area increased by about 20\%.

The optical counterpart of \sxp, the B0\,III\,e star 2dFS\,3831, was already observed spectroscopically
by \citet{2004MNRAS.353..601E} in Sep 1998 (blue) and Sep 1999 (red), the later $\sim$4 months after the ASCA observation.
On 2010 Oct 25, 2dFS\,3831 was observed spectroscopically with the VLT FLAMES instrument, only 4 days before the first optical outburst.
These spectra are discussed in \citet[][]{2012MNRAS.420L..13H}.
Our reported SALT spectrum was taken during the decline of the second outburst (dashed lines in Fig.~\ref{fig:lc}).

The post-outburst SALT spectrum (see Fig.~\ref{fig:opt_spectrum}) is similar to
those previously published and show that the Be star disc was at large not affected during the X-ray outburst, 
as it can be the case in type-II outbursts.
The hydrogen Balmer lines have contributing emission filling them in, most likely as a result of
a large circumstellar disc. This is also corroborated by a non-split shape
of the H$\alpha$ line, indicating the disc large in size and/or its
inclination is small \citep{1995A&A...302..751H}. 
However, the H$\beta$ and H$\gamma$ emission lines show split shapes.

Despite overall similarity, there are some differences
in spectra obtained at different epochs. The most apparent is the
increasing strength of the H$\gamma$ line occurring between the 2dF and the VLT measurement.
In the SALT spectrum, the H$\gamma$ emission line is still seen.
The VLT spectrum shows a stronger red-ward wing contribution, while it is more symmetrical in the SALT spectrum
pointing to V/R variability (see Fig.~\ref{fig:opt_spectrum_zoom}, {\it right}).  

Another evidence for an increasing size or density of the disc is given by the increasing 
equivalent width of the H$\alpha$ and H$\beta$ emission lines between the 2dF and the SALT measurement by ($21.0\pm0.5$)\% and ($52\pm21$)\%, respectively.
Thus, the spectroscopic measurements suggest that the decretion disc has been growing over the last $\sim$14 years. 
This may help to explain the non-detections of \sxp\ by earlier X-ray missions (Sec.~\ref{sec:X-rayLC}),
because only a small or no decretion disc might have existed in the past (at least before 1999).

The optical and X-ray spectra of \sxp\ are similar to those observed from other BeXRBs. 
The unusually slow spin period and its evolution (so far no spin-up was observed), hence, are even more surprising. 
The young long-period pulsar \sxp\ continues to challenge the present models of accreting pulsars.

\section{Conclusions}
\label{sec:conclusion}

We presented the analysis of a recent \xmm\ observation of \sxp\ and 
discuss the results in the context of complementary \swift/XRT and optical data.
The main results and conclusions are as follows:
\begin{enumerate}
 \item The NS continued to spin down to a spin period of $P_{\rm s}=(1071.01\pm0.16)$ s as measured on 2012 Oct 14. 
       This implies an average $\dot{P}_{\rm s} = (2.27\pm0.44)$ s yr$^{-1}$ during a baseline of 915 days.
 \item We do not see significant spectral changes in the X-ray spectrum, compared to the 2010 observations, 
       despite of an indication of a soft excess and Fe K$_{\alpha}$ at higher luminosities ($L_{\rm X}=3\times 10^{36}$ erg s$^{-1}$).
 \item We see indications of emission lines in the RGS spectra from prominent thermal emission lines.
       The SNR can unlikely account completely for this component.
 \item No convincing indication of proton cyclotron-absorption lines are found.  
       If we demand an absorption line with $EW>200$ eV, this would exclude a magnetic field of the NS between $4.4\times10^{13}$ G and  $1.2\times10^{15}$ G.
 \item The X-ray light curve suggests that \sxp\ is currently in a persistent X-ray emitting state with a luminosity around $L_{\rm X} =8\times 10^{35}$ erg s$^{-1}$.
       Optical and X-ray outbursts might correlate and be caused by type-I outbursts.
 \item The optical spectra obtained during the X-ray outburst are
       morphologically similar to the pre-outburst spectra obtained in 1998/99 and 2010. 
       Strong Balmer emission lines, originating from the circumstellar disc of the Be star, are present during all observations.
       This indicate that the disc is large and not strongly affected by the X-ray outburst.
 \item The evolution of the Balmer emission lines provides evidence for an increase of the   
       circumstellar disc over the last $\sim$14 years.
       This can explain the observed increase in the persistent X-ray luminosity of \sxp\ during this time. 
\end{enumerate}

\begin{acknowledgements}
We thank the \xmm\ team for scheduling the ToO observation.
The XMM-Newton project is supported by the Bundesministerium f\"ur Wirtschaft und 
Technologie/Deutsches Zentrum f\"ur Luft- und Raumfahrt (BMWI/DLR, FKZ 50 OX 0001)
and the Max-Planck Society. 
We acknowledge the use of public data from the Swift data archive 
and thank the Swift team for accepting and scheduling the ToO observation.
Some of the observations reported in this paper were obtained with the Southern African Large Telescope (SALT).
The OGLE project has received funding from the European Research Council
under the European Community's Seventh Framework Programme
(FP7/2007-2013) / ERC grant agreement no. 246678 to AU.
MPES is funded through the Claude Leon Foundation Postdoctoral Fellowship program and the National Research Foundation.
LMO and RS acknowledge support from the BMWI/DLR grants FKZ 50 OR 1302 and FKZ 50 OR 0907, respectively.
LMO acknowledges fruitful discussions within the international team at ISSI (International Space Science Institute) in Bern, especially with M.J. Torrjon and P. Kretschmar. 
JSG thanks the University of Wisconsin-Madison College of Letters \& Science for partial support of this research and the campus for funding the Wisconsin participation in SALT.
\end{acknowledgements}

\bibliographystyle{aa}
\bibliography{../auto,../general}

\begin{thebibliography}{42}
\expandafter\ifx\csname natexlab\endcsname\relax\def\natexlab#1{#1}\fi

\bibitem[{{Arnaud}(1996)}]{1996ASPC..101...17A}
{Arnaud}, K.~A. 1996, in Astronomical Society of the Pacific Conference Series,
  Vol. 101, Astronomical Data Analysis Software and Systems V, ed. G.~H.
  {Jacoby} \& J.~{Barnes}, 17

\bibitem[{{Bildsten} {et~al.}(1997){Bildsten}, {Chakrabarty}, {Chiu}, {Finger},
  {Koh}, {Nelson}, {Prince}, {Rubin}, {Scott}, {Stollberg}, {Vaughan},
  {Wilson}, \& {Wilson}}]{1997ApJS..113..367B}
{Bildsten}, L., {Chakrabarty}, D., {Chiu}, J., {et~al.} 1997, \apjs, 113, 367

\bibitem[{{Bird} {et~al.}(2012){Bird}, {Coe}, {McBride}, \&
  {Udalski}}]{2012MNRAS.423.3663B}
{Bird}, A.~J., {Coe}, M.~J., {McBride}, V.~A., \& {Udalski}, A. 2012, \mnras,
  423, 3663

\bibitem[{{Buckley} {et~al.}(2006{\natexlab{a}}){Buckley}, {Burgh}, {Cottrell},
  {Nordsieck}, {O'Donoghue}, \& {Williams}}]{2006SPIE.6269E...8B}
{Buckley}, D.~A.~H., {Burgh}, E.~B., {Cottrell}, P.~L., {et~al.}
  2006{\natexlab{a}}, in Society of Photo-Optical Instrumentation Engineers
  (SPIE) Conference Series, Vol. 6269, Society of Photo-Optical Instrumentation
  Engineers (SPIE) Conference Series

\bibitem[{{Buckley} {et~al.}(2006{\natexlab{b}}){Buckley}, {Swart}, \&
  {Meiring}}]{2006SPIE.6267E..32B}
{Buckley}, D.~A.~H., {Swart}, G.~P., \& {Meiring}, J.~G. 2006{\natexlab{b}}, in
  Society of Photo-Optical Instrumentation Engineers (SPIE) Conference Series,
  Vol. 6267, Society of Photo-Optical Instrumentation Engineers (SPIE)
  Conference Series

\bibitem[{{Burgh} {et~al.}(2003){Burgh}, {Nordsieck}, {Kobulnicky}, {Williams},
  {O'Donoghue}, {Smith}, \& {Percival}}]{2003SPIE.4841.1463B}
{Burgh}, E.~B., {Nordsieck}, K.~H., {Kobulnicky}, H.~A., {et~al.} 2003, in
  Society of Photo-Optical Instrumentation Engineers (SPIE) Conference Series,
  Vol. 4841, Society of Photo-Optical Instrumentation Engineers (SPIE)
  Conference Series, ed. M.~{Iye} \& A.~F.~M. {Moorwood}, 1463--1471

\bibitem[{{Cash}(1979)}]{1979ApJ...228..939C}
{Cash}, W. 1979, \apj, 228, 939

\bibitem[{{Crawford} {et~al.}(2010){Crawford}, {Still}, {Schellart}, {Balona},
  {Buckley}, {Dugmore}, {Gulbis}, {Kniazev}, {Kotze}, {Loaring}, {Nordsieck},
  {Pickering}, {Potter}, {Romero Colmenero}, {Vaisanen}, {Williams}, \&
  {Zietsman}}]{2010SPIE.7737E..54C}
{Crawford}, S.~M., {Still}, M., {Schellart}, P., {et~al.} 2010, in Society of
  Photo-Optical Instrumentation Engineers (SPIE) Conference Series, Vol. 7737,
  Society of Photo-Optical Instrumentation Engineers (SPIE) Conference Series

\bibitem[{{den Herder} {et~al.}(2001){den Herder}, {Brinkman}, {Kahn},
  {Branduardi-Raymont}, {Thomsen}, {Aarts}, {Audard}, {Bixler}, {den Boggende},
  {Cottam}, {Decker}, {Dubbeldam}, {Erd}, {Goulooze}, {G{\"u}del}, {Guttridge},
  {Hailey}, {Janabi}, {Kaastra}, {de Korte}, {van Leeuwen}, {Mauche},
  {McCalden}, {Mewe}, {Naber}, {Paerels}, {Peterson}, {Rasmussen}, {Rees},
  {Sakelliou}, {Sako}, {Spodek}, {Stern}, {Tamura}, {Tandy}, {de Vries},
  {Welch}, \& {Zehnder}}]{2001A&A...365L...7D}
{den Herder}, J.~W., {Brinkman}, A.~C., {Kahn}, S.~M., {et~al.} 2001, \aap,
  365, L7

\bibitem[{{Evans} {et~al.}(2004){Evans}, {Howarth}, {Irwin}, {Burnley }, \&
  {Harries}}]{2004MNRAS.353..601E}
{Evans}, C.~J., {Howarth}, I.~D., {Irwin}, M.~J., {Burnley }, A.~W., \&
  {Harries}, T.~J. 2004, \mnras, 353, 601

\bibitem[{{Fu} \& {Li}(2012)}]{2012ApJ...757..171F}
{Fu}, L. \& {Li}, X.-D. 2012, \apj, 757, 171

\bibitem[{{Gregory} \& {Loredo}(1996)}]{1996ApJ...473.1059G}
{Gregory}, P.~C. \& {Loredo}, T.~J. 1996, ApJ, 473, 1059

\bibitem[{{Haberl} {et~al.}(2008){Haberl}, {Eger}, \&
  {Pietsch}}]{2008A&A...489..327H}
{Haberl}, F., {Eger}, P., \& {Pietsch}, W. 2008, \aap, 489, 327

\bibitem[{{Haberl} {et~al.}(2000){Haberl}, {Filipovi{\'c}}, {Pietsch}, \&
  {Kahabka}}]{2000A&AS..142...41H}
{Haberl}, F., {Filipovi{\'c}}, M.~D., {Pietsch}, W., \& {Kahabka}, P. 2000,
  \aaps, 142, 41

\bibitem[{{Haberl} \& {Pietsch}(2005)}]{2005A&A...438..211H}
{Haberl}, F. \& {Pietsch}, W. 2005, \aap, 438, 211

\bibitem[{{Haberl} {et~al.}(2012){Haberl}, {Sturm}, {Filipovi{\'c}}, {Pietsch},
  \& {Crawford}}]{2012A&A...537L...1H}
{Haberl}, F., {Sturm}, R., {Filipovi{\'c}}, M.~D., {Pietsch}, W., \&
  {Crawford}, E.~J. 2012, \aap, 537, L1

\bibitem[{{H{\'e}nault-Brunet} {et~al.}(2012){H{\'e}nault-Brunet}, {Oskinova},
  {Guerrero}, {Sun}, {Chu}, {Evans}, {Gallagher}, {Gruendl}, \&
  {Reyes-Iturbide}}]{2012MNRAS.420L..13H}
{H{\'e}nault-Brunet}, V., {Oskinova}, L.~M., {Guerrero}, M.~A., {et~al.}
  2012, \mnras, 420, L13

\bibitem[{{Hickox} {et~al.}(2004){Hickox}, {Narayan}, \&
  {Kallman}}]{2004ApJ...614..881H}
{Hickox}, R.~C., {Narayan}, R., \& {Kallman}, T.~R. 2004, \apj, 614, 881

\bibitem[{{Hummel} \& {Vrancken}(1995)}]{1995A&A...302..751H}
{Hummel}, W. \& {Vrancken}, M. 1995, \aap, 302, 751

\bibitem[{{Ikhsanov}(2012)}]{2012MNRAS.424L..39I}
{Ikhsanov}, N.~R. 2012, \mnras, 424, L39

\bibitem[{{Jansen} {et~al.}(2001){Jansen}, {Lumb}, {Altieri}, {Clavel}, {Ehle},
  {Erd}, {Gabriel}, {Guainazzi}, {Gondoin}, {Much}, {Munoz}, {Santos},
  {Schartel}, {Texier}, \& {Vacanti}}]{2001A&A...365L...1J}
{Jansen}, F., {Lumb}, D., {Altieri}, B., {et~al.} 2001, \aap, 365, L1

\bibitem[{{Kobulnicky} {et~al.}(2003){Kobulnicky}, {Nordsieck}, {Burgh},
  {Smith}, {Percival}, {Williams}, \& {O'Donoghue}}]{2003SPIE.4841.1634K}
{Kobulnicky}, H.~A., {Nordsieck}, K.~H., {Burgh}, E.~B., {et~al.} 2003, in
  Society of Photo-Optical Instrumentation Engineers (SPIE) Conference Series,
  Vol. 4841, Society of Photo-Optical Instrumentation Engineers (SPIE)
  Conference Series, ed. M.~{Iye} \& A.~F.~M. {Moorwood}, 1634--1644

\bibitem[{{Mereghetti}(2008)}]{2008A&ARv..15..225M}
{Mereghetti}, S. 2008, \aapr, 15, 225

\bibitem[{{O'Donoghue} {et~al.}(2006){O'Donoghue}, {Buckley}, {Balona},
  {Bester}, {Botha}, {Brink}, {Carter}, {Charles}, {Christians}, {Ebrahim},
  {Emmerich}, {Esterhuyse}, {Evans}, {Fourie}, {Fourie}, {Gajjar}, {Gordon},
  {Gumede}, {de Kock}, {Koeslag}, {Koorts}, {Kriel}, {Marang}, {Meiring},
  {Menzies}, {Menzies}, {Metcalfe}, {Meyer}, {Nel}, {O'Connor}, {Osman}, {Du
  Plessis}, {Rall}, {Riddick}, {Romero-Colmenero}, {Potter}, {Sass},
  {Schalekamp}, {Sessions}, {Siyengo}, {Sopela}, {Stoffels}, {Scholtz},
  {Swart}, {Swat}, {Swiegers}, {Tiheli}, {Vaisanen}, {Whittaker}, \& {van
  Wyk}}]{2006MNRAS.372..151O}
{O'Donoghue}, D., {Buckley}, D.~A.~H., {Balona}, L.~A., {et~al.} 2006, \mnras,
  372, 151

\bibitem[{{Oskinova} {et~al.}(2013{\natexlab{a}}){Oskinova}, {Guerrero},
  {H{\'e}nault-Brunet}, {Sun}, {Chu}, {Evans}, {Gallagher}, {Gruendl}, \&
  {Reyes-Iturbide}}]{2013IAUS..291..459O}
{Oskinova}, L.~M., {Guerrero}, M.~A., {H{\'e}nault-Brunet}, V., {et~al.}
  2013{\natexlab{a}}, in IAU Symposium, Vol. 291, IAU Symposium, 459--461

\bibitem[{{Oskinova} {et~al.}(2013{\natexlab{b}}){Oskinova}, {Sun}, {Evans},
  {H{\'e}nault -Brunet}, {Chu}, {Gallagher}, {Guerrero}, {Gruendl},
  {G{\"u}del}, {Silich}, {Chen}, {Naz{\'e}}, {Hainich}, \&
  {Reyes-Iturbide}}]{2013ApJ...765...73O}
{Oskinova}, L.~M., {Sun}, W., {Evans}, C.~J., {et~al.} 2013{\natexlab{b}},
  \apj, 765, 73

\bibitem[{{Popov} \& {Turolla}(2012)}]{2012MNRAS.421L.127P}
{Popov}, S.~B. \& {Turolla}, R. 2012, \mnras, 421, L127

\bibitem[{{Pottschmidt} {et~al.}(2012){Pottschmidt}, {Suchy}, {Rivers},
  {Rothschild}, {Marcu}, {Barrag{\'a}n}, {K{\"u}hnel}, {F{\"u}rst}, {Schwarm},
  {Kreykenbohm}, {Wilms}, {Sch{\"o}nherr}, {Caballero}, {Camero-Arranz},
  {Bodaghee}, {Doroshenko}, {Klochkov}, {Santangelo}, {Staubert}, {Kretschmar},
  {Wilson-Hodge}, {Finger}, \& {Terada}}]{2012AIPC.1427...60P}
{Pottschmidt}, K., {Suchy}, S., {Rivers}, E., {et~al.} 2012, in American
  Institute of Physics Conference Series, Vol. 1427, American Institute of
  Physics Conference Series, ed. R.~{Petre}, K.~{Mitsuda}, \& L.~{Angelini},
  60--67

\bibitem[{{Protassov} {et~al.}(2002){Protassov}, {van Dyk}, {Connors},
  {Kashyap}, \& {Siemiginowska}}]{2002ApJ...571..545P}
{Protassov}, R., {van Dyk}, D.~A., {Connors}, A., {Kashyap}, V.~L., \&
  {Siemiginowska}, A. 2002, \apj, 571, 545

\bibitem[{{Reig}(2011)}]{2011Ap&SS.332....1R}
{Reig}, P. 2011, \apss, 332, 1

\bibitem[{{Saxton} {et~al.}(2008){Saxton}, {Read}, {Esquej}, {Freyberg},
  {Altieri}, \& {Bermejo}}]{2008A&A...480..611S}
{Saxton}, R.~D., {Read}, A.~M., {Esquej}, P., {et~al.} 2008, \aap, 480, 611

\bibitem[{{Schmidtke} {et~al.}(2012{\natexlab{a}}){Schmidtke}, {Cowley}, \&
  {Udalski}}]{2012ATel.4399....1S}
{Schmidtke}, P.~C., {Cowley}, A.~P., \& {Udalski}, A. 2012{\natexlab{a}}, The
  Astronomer's Telegram, 4399, 1

\bibitem[{{Schmidtke} {et~al.}(2012{\natexlab{b}}){Schmidtke}, {Cowley}, \&
  {Udalski}}]{2012ATel.4596....1S}
{Schmidtke}, P.~C., {Cowley}, A.~P., \& {Udalski}, A. 2012{\natexlab{b}}, The
  Astronomer's Telegram, 4596, 1

\bibitem[{{Str{\"u}der} {et~al.}(2001){Str{\"u}der}, {Briel}, {Dennerl},
  {Hartmann}, {Kendziorra}, {Meidinger}, {Pfeffermann}, {Reppin}, {Aschenbach},
  {Bornemann}, {Br{\"a}uninger}, {Burkert}, {Elender}, {Freyberg}, {Haberl},
  {Hartner}, {Heuschmann}, {Hippmann}, {Kastelic}, {Kemmer}, {Kettenring},
  {Kink}, {Krause}, {M{\"u}ller}, {Oppitz}, {Pietsch}, {Popp}, {Predehl},
  {Read}, {Stephan}, {St{\"o}tter}, {Tr{\"u}mper}, {Holl}, {Kemmer}, {Soltau},
  {St{\"o}tter}, {Weber}, {Weichert}, {von Zanthier}, {Carathanassis}, {Lutz},
  {Richter}, {Solc}, {B{\"o}ttcher}, {Kuster}, {Staubert}, {Abbey}, {Holland},
  {Turner}, {Balasini}, {Bignami}, {La Palombara}, {Villa}, {Buttler},
  {Gianini}, {Lain{\'e}}, {Lumb}, \& {Dhez}}]{2001A&A...365L..18S}
{Str{\"u}der}, L., {Briel}, U., {Dennerl}, K., {et~al.} 2001, \aap, 365, L18

\bibitem[{{Tody}(1993)}]{1993ASPC...52..173T}
{Tody}, D. 1993, in Astronomical Society of the Pacific Conference Series,
  Vol.~52, Astronomical Data Analysis Software and Systems II, ed. R.~J.
  {Hanisch}, R.~J.~V. {Brissenden}, \& J.~{Barnes}, 173

\bibitem[{{Turner} {et~al.}(2001){Turner}, {Abbey}, {Arnaud}, {Balasini},
  {Barbera}, {Belsole}, {Bennie}, {Bernard}, {Bignami}, {Boer}, {Briel},
  {Butler}, {Cara}, {Chabaud}, {Cole}, {Collura}, {Conte}, {Cros}, {Denby},
  {Dhez}, {Di Coco}, {Dowson}, {Ferrando}, {Ghizzardi}, {Gianotti}, {Goodall},
  {Gretton}, {Griffiths}, {Hainaut}, {Hochedez}, {Holland}, {Jourdain},
  {Kendziorra}, {Lagostina}, {Laine}, {La Palombara}, {Lortholary}, {Lumb},
  {Marty}, {Molendi}, {Pigot}, {Poindron}, {Pounds}, {Reeves}, {Reppin},
  {Rothenflug}, {Salvetat}, {Sauvageot}, {Schmitt}, {Sembay}, {Short},
  {Spragg}, {Stephen}, {Str{\"u}der}, {Tiengo}, {Trifoglio}, {Tr{\"u}mper},
  {Vercellone}, {Vigroux}, {Villa}, {Ward}, {Whitehead}, \&
  {Zonca}}]{2001A&A...365L..27T}
{Turner}, M.~J.~L., {Abbey}, A., {Arnaud}, M., {et~al.} 2001, \aap, 365, L27

\bibitem[{{Udalski}(2008)}]{2008AcA....58..187U}
{Udalski}, A. 2008, \actaa, 58, 187

\bibitem[{{Udalski} {et~al.}(2008){Udalski}, {Szymanski}, {Soszynski}, \&
  {Poleski}}]{2008AcA....58...69U}
{Udalski}, A., {Szymanski}, M.~K., {Soszynski}, I., \& {Poleski}, R. 2008,
  \actaa, 58, 69

\bibitem[{{Wang} \& {Wu}(1992)}]{1992ApJS...78..391W}
{Wang}, Q. \& {Wu}, X. 1992, \apjs, 78, 391

\bibitem[{{Wilms} {et~al.}(2000){Wilms}, {Allen}, \&
  {McCray}}]{2000ApJ...542..914W}
{Wilms}, J., {Allen}, A., \& {McCray}, R. 2000, \apj, 542, 914

\bibitem[{{Yokogawa} {et~al.}(2003){Yokogawa}, {Imanishi}, {Tsujimoto},
  {Koyama}, \& {Nishiuchi}}]{2003PASJ...55..161Y}
{Yokogawa}, J., {Imanishi}, K., {Tsujimoto}, M., {Koyama}, K., \& {Nishiuchi},
  M. 2003, \pasj, 55, 161

\bibitem[{{Zane} {et~al.}(2001){Zane}, {Turolla}, {Stella}, \&
  {Treves}}]{2001ApJ...560..384Z}
{Zane}, S., {Turolla}, R., {Stella}, L., \& {Treves}, A. 2001, \apj, 560, 384

\end{thebibliography}

\end{document}